\documentclass[twocolumn,aps,prl]{revtex4}

\usepackage{amsmath,amsfonts,amssymb}
\usepackage{wrapfig}
\usepackage{graphicx}
\usepackage{bbm}
\usepackage{color}

\usepackage{textcomp}
\usepackage{amssymb}
\usepackage[english]{babel}

\date{\today}

\begin{document}

\title{On machine creativity and the notion of free will}

\author{Hans J. Briegel}

\affiliation{$^1$Institut f{\"u}r Theoretische Physik,
Universit{\"a}t Innsbruck, Technikerstra{\ss }e 25, A-6020 Innsbruck, Austria\\ $^2$Institut f{\"u}r Quantenoptik und Quanteninformation der \"Osterreichischen Akademie der Wissenschaften, Innsbruck, Austria
}

\begin{abstract}
We discuss the possibility of freedom of action in embodied systems that are, with no exception and at all scales of their body, subject to physical law. We relate the discussion to a model of an artificial agent that exhibits a primitive notion of creativity and freedom in dealing with its environment, which is part of a recently introduced scheme of information processing called \emph{projective simulation}. This provides an explicit proposal on how we can reconcile our understanding of universal physical law with the idea that higher biological entities can acquire a notion of freedom that allows them to increasingly detach themselves from a strict causal embedding into the surrounding world.
\end{abstract}

\maketitle

\section{Introduction}

Are we free in our decisions and actions? Or is free will an illusion and is what we think and how we act entirely determined
by the laws of Nature?
Recent developments in brain research have revived and stirred-up a centuries-old discussion, claiming that free will is essentially an illusion \cite{SoonBrass08,Haggard08,HaynesRees06,Libet85,Wegner02}. The discussion is not only of academic nature, but it has for example been suggested that the experimental findings of the neuro-sciences, together with their theoretical interpretations, should be reflected in future jurisdiction \cite{Singer04}. These developments have lead to a controversial debate between brain researchers, philosophers, law makers, behavior scientists, and others (see e.g.\ \cite{Geyer04,MartinHeisenberg09}).

Considering what seems to be at stake, these reactions are not surprising. At the same time, they also emphasize the deep impact of the concepts and findings of modern science, in particular physics, neurobiology, and computer science, on the idea of human existence and responsibility.

The problem of free will has a long history in philosophy and science. We shall not try to give a full account of the various philosophical arguments that have been brought up against or in favor of free will. It seems however save to say that, up-to-date, the problem of free will has remained a deeply puzzling problem that many consider as yet unsolved:

\begin{quote}
\emph{``So it really does look as if everything we know about physics forces us to some form of denial of human freedom.''}\newline
--- John Searle, \emph{Minds, Brains and Science}, p. 87 (1984).
\end{quote}

This quote, out of a famous lecture by John Searle \cite{Searle84}, dates back more than 25 years, and it addresses a question of principle \footnote{In recent writing \cite{Searle07}, Searle has slightly relaxed this strong view, considering quantum mechanical indeterminism as a possible way out of this dilemma, but he is not very explicit nor optimistic about this point.}. It seems to us that this problem should be solved before any interpretation of experimental findings in the neurosciences (concerning the existence or non-existence of free will) can be reached.\newline
Indeed, how can we accept, the \emph{very possibility} of free will %(far from being able to offer a detailed theory of it)
if we assume, at the same time, that we are, with no exception and at all scales of our body, subject to physical law? Do we have to assume that the laws of physics are \emph{incomplete} and that there are new kinds of laws waiting to be discovered -- maybe on the level of more complex biological entities -- that will ultimately free us from a strict causal embedding into the evolution of the physical world?

We will argue in this note that we do not need new laws to resolve this puzzle. We may still discover new laws in the future, which will hopefully help us to better understand the workings of the human mind and all that comes with it. But I claim we shall not need such laws to resolve the conundrum of the freedom of action. We can show, on the basis of physical laws as we understand them today, that entities with a certain degree of physical or biological organization, capable of evolving a specific type of memory, can indeed \emph{detach} themselves from an strict causal embedding into the surrounding world and develop a original notion of creativity and freedom in their dealing with the environment. Our argument will be based on the idea of \emph{projective simulation} which is a new physical model of information processing for artificial agents that was recently introduced in \cite{BriegelCuevas11}.

Many philosophers and scientists have addressed the problem of free will in the past, and have argued \emph{for} the possibility of free will.
This includes, in particular, a number of theories and ideas that have been refereed to as ``two-stage models'' for free will \cite{TwoStageModels}.
At the same time, the idea of freedom is ``under attack as never before'' as was recently observed by neurobiologist Martin Heisenberg in an essay to Nature \cite{MartinHeisenberg09}. Are the experimental findings of modern brain research indeed so compelling that they could falsify all theories supporting free will?

In this paper, we would like to add another perspective to this discussion. Rather than discussing the existence of free will in the context of current brain research, which we prefer to leave to the experts, we shall present a model of an \emph{artificial agent} that exhibits a notion of freedom in dealing with its environment, which is part of a physically well-defined scheme of information processing and learning \cite{BriegelCuevas11}. This model could in principle be realized, with present-day technology, in artificial agents such as robots. This demonstrates, first, that a notion of freedom can indeed exist for entities that operate, without exception and at all scales, under the laws of physics.
It also shows that freedom of action can be understood as an emergent property of biological systems of sufficient complexity that have evolved a specific form of memory.

Formally, our proposal might be listed under the heading of the two-stage models, but it differs from previous work in several essential respects.
\begin{enumerate}
\item
We take an explicit perspective from \emph{physics} and \emph{information processing}. We introduce projective simulation as a physical concept that gives room for a notion of freedom compatible with the laws of physics.
\item
Together with the model of episodic and compositional memory, our theory of projective simulation should allow us to analyze and propose behavior experiments with simple animals.
\item
Our scheme could, in principle, be realized with present-day technology in form of artificial (learning) agents or robots.
\end{enumerate}

Finally, we want to emphasize that our model is \emph{not} meant to be an ``explanation of consciousness'' \cite{Dennett91}, nor a theory of ``how the brain works''. We leave this to the experts and to the brain researchers, and we are looking forward to the many new experimental findings and insights that we may expect to learn about in the years to come. Similarly, we are not claiming that we can explain the nature of human freedom and \emph{conscious choice}.

What we \emph{can} provide, however, is an explicit proposal on how we can reconcile our understanding of universal \emph{physical law} with the idea that higher biological entities can acquire a notion of freedom. It allows them to detach themselves from an strict causal embedding into the surrounding world and, at the same time, to truly generate behavior on their own that is both spontaneous and meaningful in response to their environment.

\section{Machine intelligence and creativity}

If we accept that free will is compatible with physical law, we also have to accept that it must be possible, in principle, to build a \emph{machine} that would exhibit similar forms of freedom as the one we usually ascribe to humans and certain animals. It is likely to turn out that the task of building such a machine will be far too complex to be realizable in any practical terms, or that it will be at least as complex as the task (and pleasure) of raising and educating a human child within society. This observation, if true, may be disappointing to some people, but for many of us it has a positive and reconciling aspect. On the other hand, it may still be feasible to build more primitive forms of machines (or agents) that exhibit some \emph{rudimentary} forms of freedom and creativity in their behavior.

\emph{Computers} are special sorts of machines which play an increasingly important role in our modern society. They have not only transformed our practical daily life, but they are also beginning to change the perception of ourselves from ``human subjects'' to ``information processing systems''. This will ultimately challenge the question of human existence and freedom, and all that comes with it (e.g.\ social responsibility, the ethics of action, and so on).
Variants of computers are  so-called \emph{intelligent agents} and robots. They are often viewed (not quite correctly, though \cite{PfeifferScheier99}) as computers equipped with some periphery, including sensors, with which they can perceive signals from the environment, and actuators, with which they can act on the environment. Intelligent agents are designed to operate autonomously in complex and changing environments, examples of which are traffic, remote space, or the internet. The design of intelligent agents, specifically for tasks such as \emph{learning}, has become a unifying agenda of various branches of artificial intelligence \cite{RusselNorvig03}.

Even if we are willing to accept that artificial agents, and computers in general, may exhibit some form of \emph{intelligence} \footnote{Intelligence of agents is usually defined as the capability of the agent to perceive and act on its environment in a way that maximize its chances of success \cite{RusselNorvig03}.},
we would hardly ascribe \emph{free will} to them. In return, we would not like ourselves to be identified with such an agent.
What is the reason for this disapproval? The main reason seems to be that the agent has a \emph{program} which determines, for a given input \footnote{In the most general case, the input may consist of the entire (percept) history of the agent.}, its next step of action. Its action is the result of an \emph{algorithm}: it is predictable and can e.g.\ be computed by some other machine.

The situation does not change fundamentally if the algorithm or program itself is not deterministic, as it is sometimes considered in computer science, invoking the notion of probabilistic Turing machines \cite{Papadimitriou94}. Even if randomized programs can sometimes increase the efficiency of certain computations, it is not clear what one should gain from such randomization in the present context. If, before, the agent was the slave of a deterministic program, it is then the slave of a random program \footnote{I owe this vivid phrase to Sandu Popescu.}. But random action is not the same as free action.

The disturbing point with both described variants is the idea and existence of the program itself. If physics is looking for the \emph{laws of Nature}, e.g. for the laws describing the way how things move and change in space and time, and of how they respond to our experimental inquiry, a more computer-science oriented approach looks for the \emph{program behind things}, including living beings. Both notions appear to be in fundamental conflict with our basic idea of freedom.

In this paper, we will however argue that the idea of being subjected to physical law does not contradict the possibility of freedom.
We will base our argument on the explicit description of an information processing scheme, which we call projective simulation \cite{BriegelCuevas11}, which could be part of the design of an artificial agent, a robot, or conceivably some biological entity. It combines the concepts of \emph{simulation}, \emph{episodic memory}, and \emph{randomness} into a common framework.

\section{Memory}

A crucial element for the possibility of freedom of action is the existence of \emph{memory}. By memory we mean any kind of organ, or physical device, that allows the agent to store and recall information about past experience. Generally speaking, memory allows the agent to relate its actions to its past.
Memory \emph{per se} is however not sufficient for the existence of free will. Elementary forms of memory exist already in simple animals (reflex-type agents), as in the roundworm \emph{Caenorhabditis elegans}, the well-studied sea slug \emph{Aplysia} \cite{KandelNobelPrize}, or the fruit fly \emph{Drosophila} \cite{HeisenbergPNAS}, and learning consists in the modification and shaping of the molecular details of their neural circuits due to experience. Nevertheless, we are hesitating to ascribe a notion of freedom to invertebrates such as C. elegans or Aplysia, whose actions remain simple reflexes to environmental stimuli.
%Upon a given stimulus, a reflex will follow immediately and %predictably. The agent remains in that sense completely attached %to the environment and its forces, almost like the motion of a %stone is subject to mechanical forces exerted on it by the %environment.

The brain of humans and higher primates is of course much more complex and much less understood. As we consider the brains of different species, moving from invertebrates to vertebrates including mammals, primates and the humans, the structure of their brains gets increasingly more sophisticated and complex. But it is always described by a network of neurons and synapses, and the basic principles of signal transmission and processing seem to be the same. The question then arises: How can an increasing complexity of a neural network lead to the emergence of a radically new feature and endow humans or higher primates, and arguably also simpler animals, with ``freedom'' in their actions?

The answer, it seems, must be sought in the increasingly complex organization of memory. A difference between the simple memory of Aplysia and the complex memory of higher vertebrates, is the appearance of different \emph{functions} of memory. Different from simple animals, a call of memory in humans and primates does not automatically lead to motor action. This means that there exists a platform on which memory content can be reinvoked, which is decoupled from immediate motor action. The evolutionary emergence of such a platform means that an agent with more complex memory can become increasingly detached from immediate response to environmental stimuli.

However, the actions of the agent still remain determined by the memory content, which itself was formed by the agent's percept history. In other words, the actions of the agent remain determined by its past, and there is no real notion of freedom. What is still missing is an element of \emph{spontaneity} in the agent's response to a given environmental situation. If C elegans is enslaved by the present stimuli, a more complex agent remains still enslaved by its past, i.e., the \emph{history} of its stimuli.
How could Nature get rid of such a time-delayed enslavery?

A possibility to break determinism is to introduce indeterminism (i.e. genuine randomness). But, as we have discussed earlier, it is not clear what the effect of randomization should be. If we adopt a computational or algorithmic view of the brain, we will not change anything. However, the effect of indeterminism depends on the nature of the processing and memory where it occurs. We will show that it can indeed have a positive effect on the agent, not in the sense of making some ``computations'' more efficient, but in the sense of introducing an element of \emph{creative variation} in its memory-driven interactions with the environment. Here it will be expedient to abandon the picture of the brain as a computer and, instead, propose a dynamic model of memory which is fully embedded in the agent's architecture and which grows as the agent interacts with the world.
%The notion of episodic and compositional memory, together with the scheme of projective simulation, can give rise to a primitive, but physically well-defined, notion of creative behavior.

In the next section, we will discuss an abstract scheme of memory processing which we call projective simulation. It operates entirely under the principles of physics but nevertheless exhibits an element of freedom in an agent's interaction with the environment. It is not clear whether this scheme is at all implemented in a real brain, but we claim that it could be realized, in principle, in artificial agents.

\section{Projective simulation}

In Ref.~\cite{BriegelCuevas11}, we considered a standard model of an artificial agent that is equipped with sensors and actuators, through which it can perceive its environment and act upon it, respectively. Internally, the agent has access to some kind of memory, which we shall describe below. Perceptual input can either lead to direct motor action (reflex-type scenario) or it first undergoes some processing (projective simulation) in the course of which it is related to memory.

The memory itself is of a specific type, which we call \emph{episodic \& compositional memory} (ECM). Its primary function is to store past experience of the agent in the form of episodes, which are (evaluated) sequences of remembered percepts and actions. Physically, ECM can be described as a \emph{stochastic network of clips}, where clips are the basic units of episodic memory \footnote{The notion of episodic memory was introduced by Endel Tulving \cite{Tulving72} in psychology, which was later adopted by cognitive neuroscience. The ``network of clips'' which we are introducing can be regarded as a rudimentary form of episodic memory within a physical toy model. It must however be emphasized that, one the one hand, our model of episodic memory is much more primitive and does e.g.\ not assume any encoding of time or the ability of dating experience. On the other hand, we go beyond the notion of memory as a mere ``storing device'' and introduce dynamic rules how episodes are processed and become part of learning scheme which we call projective simulation.}, corresponding to very short episodes (or patches of ``space-time memory'') \cite{BriegelCuevas11}.

\begin{figure}[tb]
\begin{center}
\begin{minipage}{8cm}
\includegraphics[width=8cm]{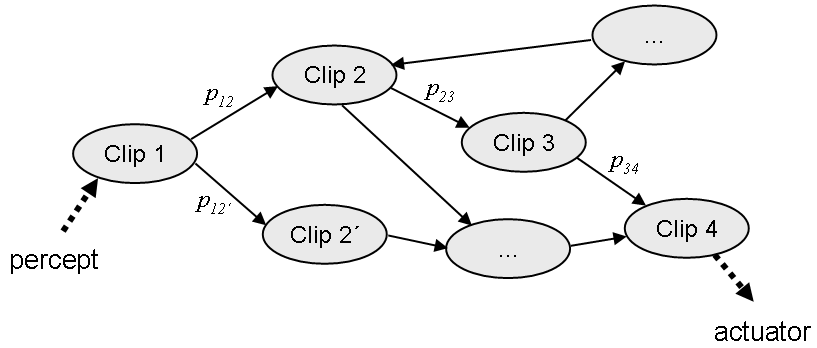}
\end{minipage}
\end{center}
\caption{Model of episodic memory as a network of clips. Triggered by perceptual input, the process of projected simulation starts a random walk through episodic memory, invoking patchwork-like sequences of virtual experience. Once a certain feature is detected,  the random walk stops and is translated into motor action (See also Ref.~\cite{BriegelCuevas11}).}
\label{ClipNetwork}
\end{figure}

The process of projective simulation can be described as follows. Triggered by perceptual input, some specific clip in memory, which relates to the input, is excited (or ``activated''), as indicated in Figure \ref{ClipNetwork}. This active clip will then, with a certain probability, excite some neighboring clip, leading to a transition within the clip network.  As the process continues, it will generate a random sequence of excited clips, which can be regarded as a recall and random reassembly of episodic fragments from the agent's past \footnote{Note that the agent does not need to be ``conscious'' for this process to be well-defined.}. This process stops once an excited clip \emph{couples out} of memory and triggers motor action. The last step could be realized by a mechanism where the excited clips are screened for the presence of certain features. When a specific feature is detected in a clip (or it is above a certain ``intensity'' level) it will, with a certain probability, lead to motor action.
\newline The decribed process is the basic version of episodic memory, but it is not the only one. In a more refined version, which we called episodic and \emph{compositional} memory, we consider not only transitions between existing clips, but clips may themselves be randomly \emph{created} (and varied), as part of the simulation process itself. Random clip sequences that are generated this way may introduce entirely fictitious episodes that never happened in the agent's past.

The random walk in memory space, as described, constitutes part of what we call \emph{projective simulation}. In another part, the agent's actions that come out of the simulation are evaluated. The result of this evaluation then feeds back into the details of the network structure of episodic memory, leading to an update of transition probabilities and of ``emotion tags'' associated with certain clip transitions \cite{BriegelCuevas11}. In a simple reinforcement setting, one assumes for example that certain actions or percept-action pairs are rewarded. Learning then takes place by modifying the network of clips (ECM) according to the given rewards.
This modification of memory occurs in different ways:
by bayesian updating of transition probabilities between existing clips; by adding new clips to the network via new perceptual input; by creating new clips from existing ones under certain compositional and variational principles; and by updating emotional tags associated with certain clip transitions.
Details of this scheme are given in Ref.~\cite{BriegelCuevas11}.

\begin{figure}[tb]
\begin{center}
\begin{minipage}{9cm}
\includegraphics[width=9cm]{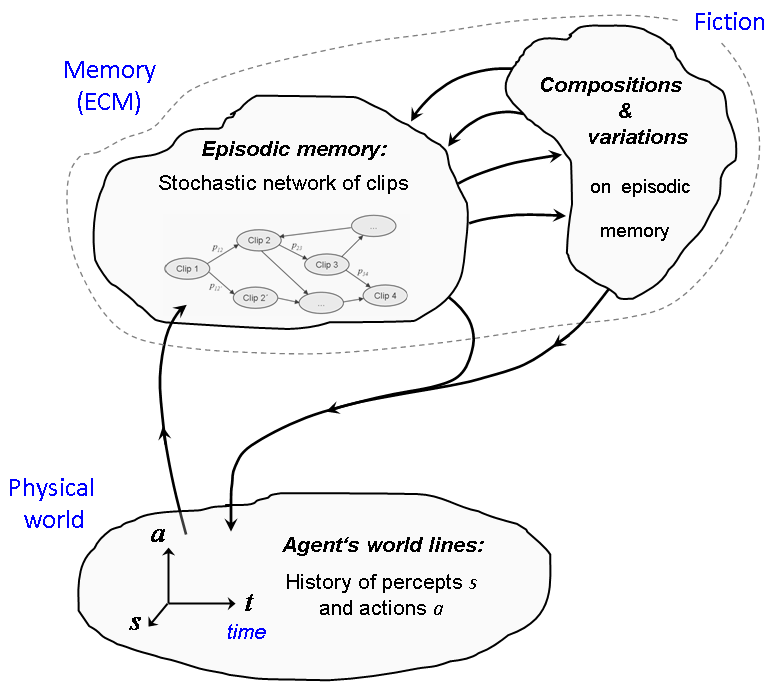}
\end{minipage}
\end{center}
\caption{Sequences of percepts and actions are simulated stochastically by variations and compositions of episodic memory (ECM), before real action is taken. Through the process of projective simulation, the agent is, in a sense, constantly ahead
of itself.}
\label{WorldMemoryFiction}
\end{figure}

In the following, coming back to the main topic of this paper, we want to relate the projective structure of the agent's behavior to the emergence of a primitive notion of creativity and freedom. The basic idea is that the episodic memory provides a platform for the agent to \emph{``play with''} previous experience, before concrete action is taken (see Figure \ref{WorldMemoryFiction}). A call of episodic memory initiates a random walk through memory space, invoking patchwork-like sequences of past experience. This can be understood as a simulation of plausible future experience on the basis of past experience.
It is a \emph{simulation} because it takes place only in the agent's memory; it simulates \emph{plausible} future experience because sequences of episodes that occurred frequently in the past will do so in simulation.
Furthermore, the possibility of clip composition allows the agent to explore, as part of the simulation, new \emph{fictitious} episodic sequences that it has never encountered before, but which are within a range of ``conceivability'' (as defined by the rules of clip composition).
It is important to realize that, in a similar way as clips representing ``real'' experience, the clips representing ``fictitious'' experience will trigger factual action through the same mechanism. This means that fictitious experience, created within the memory of the agent, may de facto change and guide the real actions of the agent. One could also say that the agent acts under the influence of ``ideas'' that are generated by the agent itself.

In summary, through the process of projected simulation the agent projects itself into conceivable future situations and takes its actions under the influence of these projections, as it is illustrated in Figure~\ref{WorldMemoryFiction}. In this sense, the agent is no longer enslaved by its past, but plays with it, deliberated by variations and spontaneous compositions of  episodic fragments. These fragments may come from the past, but they are transformed, by random processes, into new patterns for future action. The agent is, in this sense, always \emph{ahead} of itself (see discussion in the next section).

\section{Discussion}

In this section, we want to put the presented scheme into a broader philosophical context and discuss its relation to the problem of free will.

The problem of free will is often discussed in the context of conscious human experience \cite{Searle07}, specifically when we experience the freedom of choice between different options, say, of choosing between different meals in an restaurant. The problem then consists in an apparent inconsistency between such conscious experience of freedom, on the one hand, and the assumption that all of our conscious experiences are ultimately determined by neurobiological processes in the brain and as such subject to the laws of physics and biochemistry.

Other scientists see the problem of freedom already arise, perhaps in a more rudimentary form, on the level of creatures that may not be conscious, but to which we would nevertheless ascribe a measure of \emph{initiative} and self-determination in their behavior. As Martin Heisenberg puts it in Ref.~\cite{MartinHeisenberg09}: \emph{``Some define freedom as the ability to consciously decide how to act. I maintain that we need not be conscious of our decision making to be free. What matters is that our actions are self generated.''}

Whatever definition one chooses, both notions of freedom, be it in the sense of conscious free choice or in the sense of self-generated (and rational) action, have to be reconciled with the basic assumption that biological agents - conscious or unconscious - are, without exception and at all scales of their bodies, subject to physical law. The fundamental problem is, in both cases, how genuine notions of freedom can emerge from lawful processes. Both the freedom of self-generated action and the freedom of conscious choice require, at a certain level (e.g. in the brain), some notion of \emph{room to manoeuvre} \cite{Searle07}, which is consistent with physical law. Where does this room come from? And how can it be realized within an explicit physical model?

In this paper, we have discussed a model of an artificial agent, where such room for manoeuvre \footnote{In German one uses the word \emph{Spielraum} which, literally translated, means \emph{room to play}.} is provided by a specific notion of memory (ECM) and the way how this memory is used via projected simulation of future actions. Room and ultimately freedom arises in two ways, first by the existence of a simulation platform, which enables the agent to detach itself from an immediate causal embedding into its environment and, second, by the constitutive processes of the simulation, which generate a space of possibilities for responding to environmental stimuli. The mechanisms that allow the agent to \emph{explore} this space of possibilities are based on \emph{random processes}. The concept of projective simulation thus combines the basic notions of memory, randomness, and simulation in a unique way. In the remainder of this paper, we would like to come back to these ingredients of our scheme and comment on their specific role regarding the origin of freedom.

\emph{Memory.} The existence of memory is required in a trivial sense, as a physical notion of experience. But memory also provides the first step to deliberate a system from its environment, i.e.\ from an immediate stimulus-reflex-type embedding into the world. %\footnote{Deliberation from the environment does not mean a simple ``encapsulation'' or isolation of the agent, by cutting-off any interaction with the environment. The agent still remains in contact with and reacts upon the environment, but this reaction is no longer an immediate response to a stimulus but takes into account what has happened to the agent in the past.}.
By connecting perceptual input with memory content, the agent is able to relate it to past experience, on the basis of which it finds its next step. This endows the system with a comprehensive way of responding to environmental input, but its responses are still fully determined by past experience. In the specific context of episodic memory this means that, as long as episodes are simply recalled without further modification, the agent remains \emph{caught} by its past and will simply repeat old patterns of behavior. What is still missing, is the notion of the \emph{new}.

The seed of the new is provided by introducing elements of variation and composition into the simulation process. The first kind of variation is provided by a random reshuffling of past episodes, realized by a random walk in clip space. While this will already lead to new patterns of behavior, the space of possibilities is still defined by past experience alone. The second kind of variation is based on clip composition, which is able to create new fictitious episodes. It is important to realize that these variations are truly generated by the agent itself. The connection of the agent with its own past is thereby loosened and the agent becomes further ``emancipated'' from the environment. However, the agent's connection with the past is not simply blurred or erased, as it would be the case for an arbitrary randomization of memory. This would be a silly form of emancipation, depriving the system of what it may have learnt before. Instead, the agent still makes use of past experience, but it is no longer caught or enslaved by it. It rather ``plays'' with its experience in a constructive sense, creating fictitious sequences of action to guide its future actions. This type of simulation process is \emph{conservative} in the sense that only variations around real (and proven) experience are considered. It is the range of those variations that defines the conceivable. The probability of variations is determined by certain rules of clip composition, i.e.\ how memory content can mutate or, more generally, transform during the simulation process \footnote{In the language of Kant one might say that the rules of clip variation and composition define the \emph{a priori} conditions of free behavior.}. It is a stochastic process that originates and operates entirely within the memory system. In this sense, the deliberation of the agent is truly self-generated and, as such, represents a step of emancipation from its surrounding world.

\emph{Randomness.}
The notions of indeterminism and randomness play an important role in our discussion. Random processes have been assumed as part of projective simulation, both in basic memory recall --a random walk through a space of episodes-- and in the mutation or compositional processes of individual clips \footnote{Note that, when we speak of a random process with different outcomes, we mean that the different outcomes are not determined, but they occur only with certain probabilities, such as 0.1/0.9 or 0.5/0.5. We do not imply that these probabilities are all equal. Some people would instead speak of a stochastic process.}. The reader may wonder how we can postulate random processes as part of our physical model. However, this is in fact a very natural assumption, which is in agreement with the fundamental (i.e.\ quantum mechanical) laws of nature. Truly random processes are generated routinely in modern quantum physics laboratories, e.g. for quantum information processing purposes. But also in biological systems, random processes are omnipresent, a fact which has recently been emphasized in Ref.~\cite{MartinHeisenberg09}. In neural networks in the brain, for example, fluctuations in the electric potential across the neuron membrane are caused by fluctuations in the number of ions that cross the ion channels. These result from the interaction of the ions with the molecular potential of channel protein, which is described by quantum mechanics. This means that quantum mechanical indeterminism, which is a fundamental and irreducible property of molecular systems interacting with an environment, leads to fluctuations and random noise in the entire network activity. Although, for practical considerations, the origin of noise is usually not important, it is here a matter of principle.
In other words: We may not \emph{need} quantum mechanics to understand the principles of projective simulation, but we \emph{have} it. And this is our safeguard that ensures true indeterminism on a molecular level, which is amplified to random noise on a higher level. Quantum randomness is truly irreducible and provides the seed for genuine spontaneity \footnote{These observations do not exclude the possibility the quantum mechanics could even play a more constitutive role in the brain, as some people have argued, beyond the role of providing irreducible randomness. For example, in the present context, the efficiency of projected simulation might be enhanced significantly by exploiting effects of quantum coherence and interference in random walks \cite{BriegelCuevas11}. While this provides interesting new perspectives for the development of agents based on quantum technologies, it is not an essential aspect for our discussion of (the physical origins of) freedom.}.

One should also realize that the question of principle of the \emph{possibility of free will} on the basis of natural law does not depend on specifics of neurobiology. Even if people doubt the relevance of quantum indeterminacy in biological agents, they must face the possibility that sooner or later mankind may \emph{build} \emph{artificial} intelligent agents that will use quantum elements as part of their design. To put it provocatively, even if human freedom were to be an illusion, humans would still be able, in principle, to build \emph{free robots}. Amusing.

Finally, one might ask why randomness in our model of projective simulation is different from randomness in any other computational model, e.g. a boolean circuit. Why is it any better to be the slave of a random ``mutation of clips'' than of some ``randomized algorithm''? The answer is that, in the model of projective simulation, there is a clear functional role of randomness, which introduces variations around established patterns of behavior. It is on the background of previous experience, where variations \emph{make proper sense} and allow the agent to explore \emph{new possibilities} via simulation, i.e. before trying them out. This is not a notion of slavery but of self-generated choice. In contrast, randomization of an algorithm given e.g.\ in form of some  Boolean circuit may completely change the meaning and function of the algorithm and thus introduce truly \emph{meaningless} variations \footnote{This is not withstanding the fact that random noise, e.g. on the level of individual neurons in a neural network, may indeed lead to an increase of efficiency of the network in certain cases. This phenomenon is known as stochastic resonance, which is a generic feature of many nonlinear systems. It would be interesting to study a neural-network representation (in the sense of ``machine code'') of our model of episodic and compositional memory. By imposing the condition that random noise on the level of individual neurons must translate into semantically meaningful processes like clip mutation on the level of clip space, this may even give us some insight how to identify a truly computational model for neutral networks and how this can be supported by a corresponding functional organizations
of neural networks.}.
It is thus not randomness \emph{per se} that leads to deliberation or freedom. It is the (room for) variation relative to past experience that establishes a meaningful notion of choice.

\emph{Simulation.}
The physical process of simulation, combining randomness and episodic memory to generate ``virtual experience'', results in a \emph{projective structure} of the agent's behavior in its interaction with the world, as illustrated in Figure~\ref{WorldMemoryFiction}. The agent takes actions under the influence of its own projections and is, in this sense, constantly ahead of itself. It is worth pointing out that this resembles a fundamental structure (the existentiale \emph{Entwurf}) in Heidegger's phenomenology \cite{Heidegger26}, which plays an constitutive role for the notion of human understanding and \emph{being-in-the-world}. The latter provides an interesting but different framework for the discussion of human freedom, which lies much beyond the scope of the current paper. Clearly, in the present discussion we are not talking about conscious agents nor about any deeper aspect of human existence. What is remarkable, however, is that one of the fundamental structures of phenomenology can be brought in close connection with basic notions of modern physics and science \cite{BriegelUnpublished}.
It seems to us that a careful analysis of human (and animal) behavior, both from the perspective of phenomenology and of developmental psychology \cite{Tomasello99}, may indeed offer new ideas towards a better understanding of artificial intelligence \footnote{Heidegger's philosophy has influenced a number of writers including e.g.\ Terry Winograd \& Fernando Flores \cite{WinogradFlores86}, who emphasized the potential of phenomenology for new designs in computer science, and Hubert Dreyfus \cite{Dreyfus92}, who raised fundamental criticism against traditional approaches to artificial intelligence. These early works have, in turn, influenced more recent approaches to artificial intelligence, which go under the heading of embodied cognitive science (for a review see, for example, Ref.~\cite{PfeifferScheier99}).} and the ultimate possibilities of ``information processing'' in biological agents.

\end{document}